\documentclass[11pt]{article}
\usepackage[left=1in,top=1in,right=1in,bottom=1in,nohead]{geometry}
\usepackage{amsmath,amssymb,comment}
\usepackage{mathtools}
\usepackage{rotating}
\usepackage{graphicx}

\usepackage{subcaption}
\usepackage{float}
\usepackage{amsmath,amssymb}
\usepackage{amssymb}
\usepackage{booktabs}
\usepackage{amsfonts}
\usepackage{graphics,adjustbox}
\usepackage{bm}                % Bold font in formula
\usepackage{mathrsfs}      %HUA TI
\usepackage{amsmath, amsthm}
\usepackage{graphicx,epsf,pstricks,pst-node,psfrag,amsthm,amssymb,amsmath}
\usepackage{float}
\usepackage{array}
\usepackage{authblk}
\usepackage{natbib}
\usepackage{url}
\usepackage{hyperref}
\usepackage{rotating}
\usepackage{multicol}
\usepackage{multirow}
\usepackage[title]{appendix}
\usepackage{algorithmic, algorithm}

\newtheorem{theorem}{Theorem}

\newtheorem{lemma}[theorem]{Lemma}

\newcommand{\calT}{\mathit{T}}
\usepackage[normalem]{ulem}
\definecolor{orange}{rgb}{1,0.5,0}
\def\jt [#1]#2{\sout{#1} \textcolor{orange}{#2}} %track changes
\definecolor{green}{rgb}{0.513,0.73,0.442}
\def\dt [#1]#2{\sout{#1} \textcolor{green}{#2}} %track changes
\usepackage{tikz}
\usetikzlibrary{calc,shapes,bayesnet}

\newcommand{\s}{{\bf s}}

\newcommand{\y}{{\bf y}}

\newcommand{\J}{{\bf J}}

\newcommand{\blambda}{\boldsymbol{\lambda}}

\newcommand{\bTheta}{\boldsymbol{\Theta}}

\newcommand{\bgamma}{\mathbf{\gamma}}

\newcommand{\ben}{\begin{enumerate}}
\newcommand{\een}{\end{enumerate}}
\newcommand{\beq}{\begin{equation}}
\newcommand{\eeq}{\end{equation}}

\newcommand{\half}{\frac{1}{2}}

\begin{document}
	\title{Nonparametric Bayes multiresolution testing for high-dimensional rare events}
	\date{} 
	\author[1]{Jyotishka Datta}
	\author[2]{Sayantan Banerjee}
	\author[3]{David B. Dunson}
	
	\affil[1]{Virginia Tech, USA}  
	\affil[2]{IIM Indore, India}
	\affil[3]{Duke University, USA}
	
	\maketitle
	
	%\author{
		%\name{Jyotishka Datta\textsuperscript{a}\thanks{CONTACT J. DATTA. Email: jyotishka@vt.edu}, Sayantan Banerjee\textsuperscript{b} \thanks{CONTACT S. BANERJEE. Email: sayantanb@iimidr.ac.in}, David B. Dunson\textsuperscript{c}\thanks{CONTACT D. DUNSON. Email: dunson@duke.edu.}}
		%\affil{\textsuperscript{a}Virginia Tech; \textsuperscript{b}Indian Institute of Management Indore; \textsuperscript{c} Duke University}
		%}
	
	\maketitle
	
	\begin{abstract}
		In a variety of application areas, there is interest in assessing evidence of differences in the intensity of event realizations between groups. For example, in cancer genomic studies collecting data on rare variants, the focus is on assessing whether and how the variant profile changes with the disease subtype.  Motivated by this application, we develop multiresolution nonparametric Bayes tests for differential mutation rates across groups. The multiresolution approach yields fast and accurate detection of spatial clusters of rare variants, and our nonparametric Bayes framework provides great flexibility for modeling the intensities of rare variants. Some theoretical properties are also assessed, including weak consistency of our Dirichlet Process-Poisson-Gamma mixture over multiple resolutions. Simulation studies illustrate excellent small sample properties relative to competitors, and we apply the method to detect rare variants related to common variable immunodeficiency from whole exome sequencing data on 215 patients and over 60,027 control subjects.
	\end{abstract}
	
	%\begin{keywords}
	%Multiresolution testing; Nonparametric Bayes; rare events; weak consistency.
	%\end{keywords}
	
	\section{Introduction}
	\label{sec:intro}
	Rare variants, defined as `alternative forms of a gene that are present with a minor allele frequency (MAF) of less than 1\%', play a critical role in explaining the genetic contribution to complex diseases by accounting for disease risk and trait variability, previously unexplained by large genome-wide association studies focused on common variants \citep{pritchard2001rare}. In spite of the advent of low-cost parallel sequencing approaches and the resultant development of statistical and machine learning methods for rare variants (see \cite{nicolae2016association}, \cite{meng2020novel} and references therein), very few previous works fully characterize the spatial nature of the rare mutations while retaining robustness, power, and scalability to massive dimensions. In this paper, we develop a flexible, multi-scale method for assessing differences in rare mutation rates between groups by probabilistically modeling them as rare events across the whole genome. \par
	
	\textbf{Motivation} The failure of single variant tests for whole-genome sequencing data has motivated a variety of rare variant association tests, ranging from burden tests \citep{morgenthaler2007strategy, li2008methods} and adaptive burden tests \citep{lin2011general} to variance component tests, e.g. SKAT (sequence kernel association test) \citep{wu2011rare}. Unfortunately, these approaches are computationally intractable for ultra-high dimensional problems, and they ignore the physical locations on the genome. An example of strong spatial clustering of mutations is shown in Fig. \ref{fig:pik3ca}, taken from the tumor portal, on a known oncogene PIK3CA  \citep{shayesteh1999pik3ca}. To address the localization of mutations within small genomic windows, \citet{fier2012location} used an adaptive weighting scheme, and \citet{ionita2012scan} used scan-statistics \citep{kulldorff1997spatial} to find the maximum density region (MDR) for rare variants. Although scan-statistics incorporate spatial information, the fixed rate assumption is restrictive, and it allows for a single MDR, losing power in the absence of strong clustering and becoming computationally expensive. The computational complexity is $O(N \log N)$ for $N$ locations (assuming fixed cost for density evaluation of intervals), which is intractable for whole genome or whole exome data sets, worsened by the additional computational burden of permutation P-values. 
	
	Computational costs are a crucial consideration for analyzing these massive datasets.  There are $35,192,888$ possible locations on a single chromosome 22 where a rare variant could be present, out of which 312,010 positions exhibit $\mbox{mean allelic frequency} \leq 5\%$ which is by far the smallest data set available on the \emph{1000 Genomes Project}. Even the whole-exome sequencing data, representing less than 2\% of the genome, is too huge for a linear-time algorithm to handle. For example, the whole-exome sequencing data from \emph{Exome Aggregation Consortium} that forms our control group in Section~\ref{sec:real} has $6,053,186$ amino acid positions on $18,774$ different genes that could harbor a rare variant. There is a pressing need for scalable methods that incorporate spatial information, exploit sparsity to scale to huge dimensions (computationally and statistically), and appropriately characterize uncertainty in mutation rates.
	
	\begin{figure*}[ht!]
		%\vskip 0.2in
		\begin{center}
			\centerline{\includegraphics[width=0.9\linewidth,height=2in]{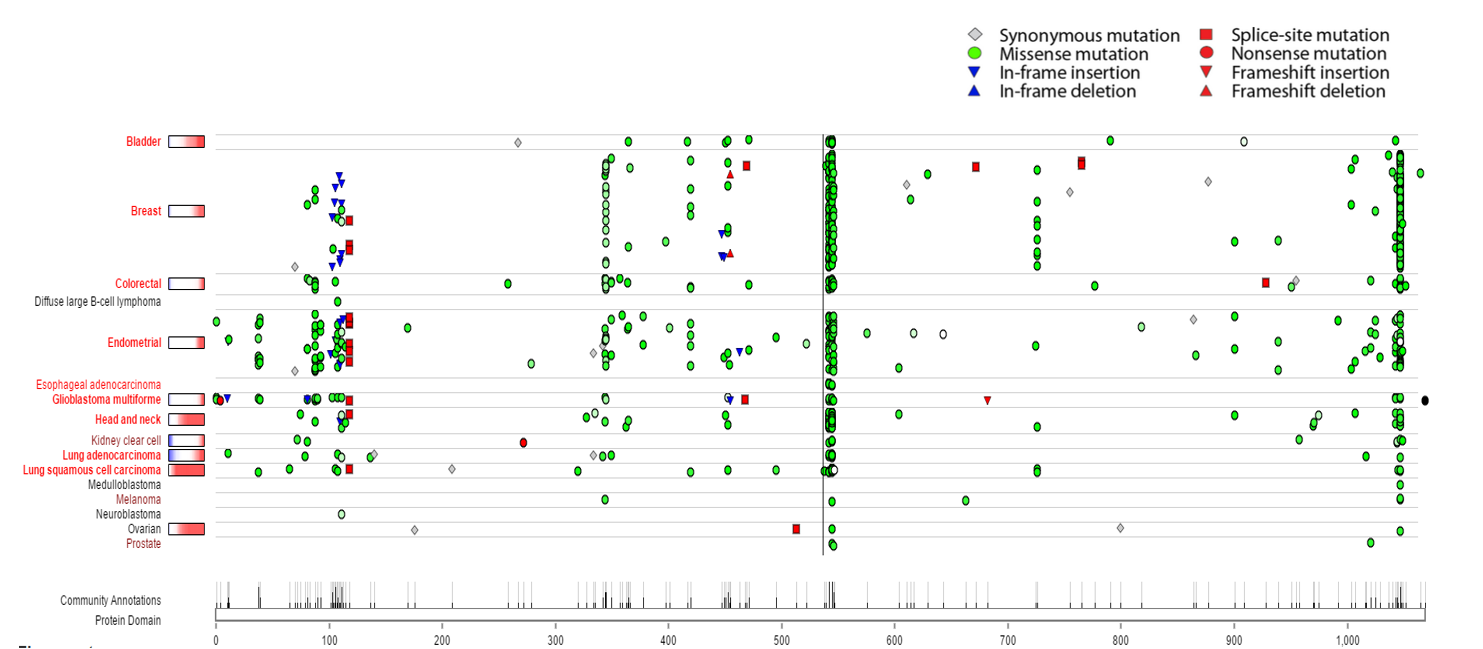}}
			\caption{Spatial Clustering on PIK3CA}
			\label{fig:pik3ca}
		\end{center}
		%\vskip -0.2in
	\end{figure*}
	
	These considerations have led us to model the rare variants as realizations of a multiscale sparse point process that achieves a) computational speed by exploiting the sparsity,  b) clustering of rare variants in smaller windows of a larger genomic region, and c) flexibility in modeling the unknown rate function. The sparsity and the clustering phenomenon of mutations imply that the intensity function will be effectively zero on a large majority of its domain and moderately higher in a few `hot spots'. We achieve computational efficiency by targeting the search to avoid genomic deserts containing no variants. We propose a binary tree-based, top-down multiresolution approach, in which we traverse from the coarsest resolution (whole genome) to finer resolutions, pruning irrelevant segments. The huge computational advantage stems from the fact that the number of expansion levels the tree must have to detect all clusters is on the order $O(\log N)$ (assuming fixed cost for density evaluation of intervals), which leads to a huge gain for very large $N$. Alternative multiresolution approaches have been used to detect high-density regions in applications including epidemiology \citep{neill2004detecting}, astronomy \citep{van2020multi}, and medical imaging \citep{neill2004fast}, having better performance than scan statistics \citep{kulldorff1997spatial} or grid-based hierarchical clustering \citep{agrawal1998automatic}. 
	
	Building a flexible model for the intensity function is another key consideration as current parametric approaches for modeling sparse count data across different conditions, primarily based on zero-inflated Poisson (see \cite{nie2006integrated, dhavala2010bayesian}) or zero-inflated negative binomial models (see \cite{robinson2010edger, hardcastle2010bayseq, anders2010differential}), do not offer sufficient flexibility in modeling the unknown mutation rate profile. The method proposed by \citet{robinson2010edger}, who model the variance as a function of both the mean and an additional component for incorporating the over-dispersion common to genomic data, is implemented in the popular R-package \textit{EdgeR}. Other approaches involve Poisson-lognormal model \citep{sepulveda2010estimation, rempala2011model}, truncated Poisson-Gamma model \citep{thygesen2006modeling}, and finite mixture of Poissons with a known number of components \citep{zuyderduyn2007statistical}. 
	Despite their popularity and success in genomic applications of relatively smaller scale, the Poisson model does not offer sufficient flexibility to model the number of variants in a given segment of the genome, in the sense that the Poisson mean also determines the rate and probability of zero variants. Instead, we propose a flexible nonparametric model that assumes that the variant frequencies are sampled from an infinite mixture of Poisson distributions that also allows for over-dispersion. Such mixture models naturally introduce latent parameters $\lambda_i$ that can be regarded as the true intensity at the $i^{th}$ location. Related non-parametric Poisson mixtures have been used in many different contexts \citep{trippa2011false}. 
	
	Our nonparametric Bayes modeling through a Dirichlet process prior on the Poisson rate parameters allows for great flexibility in modeling and performs well in estimating any true process $P_0$ in simulation studies (\textit{vide} Section~\ref{sec:sim}) and in practice identifies `differentially mutated' regions from whole-exome sequencing data for $270$ patients with common-variable immunodeficiency symptom where the background mutation rate is obtained from a publicly available \textit{ExAC} database on 60,276 individuals (\textit{vide} Section~\ref{sec:real}). Theoretically, our DPM-Poisson-Gamma model leads to posterior consistency in estimating the marginal probability mass function of the rare variants, and intuitively, the consistency will continue to hold even after combining the estimates over multiple intervals (\textit{vide} Section~\ref{sec:theory}). 
	
	%%%%%%%
	\section{Modeling Framework}\label{sec:model} 
	Our strategy is to partition the chromosome and prune `uninteresting' intervals recursively. The key insight is that whole genome data admit a natural multiscale representation from the coarsest level (whole genome) to the finest level (individual variants). There is a rich literature on multiresolution methods that provide algorithms for multiscale decomposition of observed processes and have led to excellent tools for image denoising, classification, and data compression but have been under-explored in Genomics. 
	
	We employ a nonparametric approach of modeling the number of rare variants in each sub-interval of the genome induced by a space-partitioning balanced binary tree. The nonparametric model yields great flexibility and increasing power even in the case of weak clustering of variants. The multiresolution approach also allows for multiple clusters and speeds up the computation by a top-down pruning algorithm, as we describe below. 
	
	We can imagine the mutations as outcomes of a sparse point process across the genome, with the rate function potentially varying across groups. The baseline intensity rate is zero or extremely close to zero in an overwhelming majority of the locations on the genome (`cold spots') and substantially higher in biologically interesting segments of the genome (`hot spots'). The differences between the groups manifest as shifts in the baseline rate functions that would be zero or very close to zero at the vast majority of the locations. Our goal is to calculate the posterior probability that the shift is everywhere zero (global null hypothesis) and the posterior probabilities for a sequence of increasingly local null hypotheses specific to genes or regions. 
	
	Formally, let $\calT$ be the entire region of interest, which in our case would be the whole genome or whole exome sequence under consideration. Let $\calT_p^{(l)}$, for all $p = 1, \ldots, 2^l$, for all $l = 1, \ldots, L$ denote the partition of $\calT$ at resolution $l$, where $L$ denotes the total number of resolutions. Let $S_p^{(l)} = \{ i : t_i \in \calT_p^{(l)} \}$ denote the indices of the events in the interval $\calT_p^{(l)}$, and let $n_p^{(l)} = |S_p^{(l)} |$ denote the cardinality of $S_p^{(l)}$. Let $\hat{\calT}_p^{(l)}$ denote $K$ equispaced points in the interval $\calT_p^{(l)}$, and let $y_{p,i}^{(l,1)}$ and $y_{p,i}^{(l,2)}$ denote the number of events for the two groups in the $i^{th}$ sub-interval induced by $\hat{\calT}_p^{(l)},\, i = 1, \ldots, K+1$. Our goal is to test the differences in the underlying rates for the two sequences $\y_{p}^{(l,1)}$ and $\y_{p}^{(l,2)}$ for each $p$ and $l$. 
	
	We use the following semi-parametric model proposed by \citet{guindani2014bayesian} for analyzing sequence count data on T-cell diversity for testing differential abundance (without the multi-resolution screening). Ignoring the resolution-specific subscripts $p$ and $l$ for the sake of notational clarity for the moment, let the true abundance rates for the two groups be $\lambda_{i}^{(1)}$ and $\lambda_{i}^{(2)}$. Then, the model for differential abundance is: 
	\begin{align}
		& y_{i}^{(1)} \sim  \mathrm{Poi}(\lambda_{i}^{(1)}) \mbox{ and } y_{i}^{(2)} \sim \mathrm{Poi}(\lambda_{i}^{(2)}) \nonumber \\
		& [ \lambda_{i}^{(1)} \mid P ] \stackrel{iid}{\sim} P \nonumber \\
		& [ \lambda_{i}^{(2)} \mid \lambda_{i}^{(1)}, P ] \stackrel{iid}{\sim} \pi I(\lambda_{i}^{(1)} = \lambda_{i}^{(2)}) + (1-\pi) P \; P \sim DP(M, P_{0}), \label{eq:spike}
	\end{align}
	where $M = M^{(l)}$ is the precision parameter of the Dirichlet process $P=P^{(l)}$, and $P_0 = P_{0,p}^{(l)}$ is the base measure, which is taken to be $P_{0,p}^{(l)} = \mathrm{Gam}(\alpha_p^{(l)}, \beta_p^{(l)})$. Here $\alpha_p^{(l)}, \beta_p^{(l)}$ respectively denote the resolution-specific shape and rate parameter of the Gamma distribution, which can be estimated from the data using an Empirical Bayes approach following the sequential approach of \citet{mcauliffe2006nonparametric}. The unique draws from the posterior in the inference phase are used to estimate the hyper-parameters of the Gamma base measure for the DP mixture model. A schematic diagram of the model is given in Figure \ref{fig:model}. To facilitate posterior computation, we introduce a latent indicator variable $\bgamma$ such that 
	\begin{equation*}
		\gamma_i = \begin{cases} 0 & \mbox{if } \lambda_{i}^{(1)} = \lambda_{i}^{(2)} \\
			1 & \mbox{ otherwise, } \end{cases}
	\end{equation*}
	where $\gamma_i \sim \mbox{Bern}(\pi)$. The posterior probability of differential abundance is given by: 
	\beq
	\omega_{p,i}^{(l)} = P \left(\lambda_{p,i}^{(l,2)} \neq \lambda_{i}^{(l,1)} \mid y_{i}^{(1)}, y_{i}^{(2)} \right). \label{eq:omega}
	\eeq
	To implement the Gibbs sampler, we can integrate out the base distribution $P_0$ and write the marginal likelihood of $\y$ as a negative binomial likelihood given the cluster configuration indices $s_{i,j}$ such that $s_{i,j} = k$ if $\lambda_{i,j} = \lambda_j^*$ as:
	\beq
	p(\y \mid \s) = \frac{1}{\prod_{i =1}^{K+1}y_{i,j}!} \frac{\beta^{\alpha J}}{\Gamma(\alpha)^J} \prod_{j = 1}^{J} \frac{\Gamma(\alpha+\tilde{y_j})}{(\beta+N_j)^{\alpha+\tilde{y_j}}} \label{eq:py},
	\eeq
	where $\tilde{y_j}$ and $N_j$ denotes the sum of $y_{i,j}$'s belonging to $j^{th}$ cluster and its cardinality with $J$ being the total number of clusters. To update the cluster configurations $s_{i,j}$s, we first note that $s_{i,2} = s_{i,1}$ if $\gamma_i = 0$, and $s_{i,j}$ follows a multinomial distribution with the probability vector given by: 
	\beq 
	P(s_{i,j} = k \mid \gamma_i = 1, \mbox{rest}) = \begin{cases} \frac{M}{M+\sum \gamma_i} p(\y \mid \s) \mbox{ for } k = J^{-i}+1 \\
		\frac{M}{M+\sum^{\sim k}\gamma_i} p(\y \mid \s) \mbox{ for } k \leq J^{-i}, \end{cases}
	\nonumber
	\eeq
	where $`\mbox{rest}$' denotes the remaining parameters $[\gamma_{-i}, s_{-i,j},\y$], the symbol $\sum^{\sim k}$ denotes summation over all $s_{i,j} = k$. Furthermore, $p(\y| \s)$ is given by \eqref{eq:py}, with $\tilde{y_j}$ and $N_j$ replaced by $\tilde{y_j}^{-}$ and $N_j^{-}$, the sum of counts and size of the $j^{th}$ cluster with the $i^{th}$ observation deleted, and $J^{-i}$ denotes the maximum of $\{ s_{-i,-j} \}$.  On the other hand, the full conditional distribution for the differential abundance $\gamma_i$ can be derived as follows: 
	$\gamma_i$ will be $1$ if $s_{i,2} \neq s_{i,1}$, and will follow a Bernoulli distribution with probability: 
	$p(\gamma_i \mid s_{i,1} = s_{i,2}, \mbox{rest}) = p(s_{i,2} \mid \gamma_i, s_{i,1} = s_{i,2}, \mbox{rest}) \times p(\gamma_i)$. 
	
	We can carry out a multiple testing procedure for the $K+1$ paired observations in the $(p,l)^{th}$ sub-interval with a thresholding rule on $\omega_{p,i}^{(l)}$ as follows: 
	\beq
	\mbox{Reject } H_{0,p}^{(i,l)}: \{ \lambda_{p,i}^{(l,2)} = \lambda_{i}^{(l,1)} \} \mbox{ if } \omega_{p,i}^{(l)}> \xi, \label{eq:rule}
	\eeq 
	for some suitable threshold $\xi$. For analyzing sequence count data on T-cell diversity, \citet{guindani2014bayesian} proposed a related nonparametric method for testing differential abundance (without the multi-resolution screening), that was shown to be more powerful than existing parametric procedures. 
	
	This gives us a way of pruning intervals that cannot possibly contain any differentially mutated sub-regions since the shift in the mutation rate across the groups would be essentially zero at the vast majority of the locations. We do this by calculating the probability of a `global null' hypothesis $H_{0,p}^{(l)}$ restricted to each partition $\calT_p^{(l)}$, $p = 1, \ldots, 2^l$ and pruning the interval if it exceeds a predetermined threshold. 
	
	In particular, we test the following `global' hypothesis within each partition:
	\begin{itemize}
		\item $H_{0,p}^{(l)}$: there is no difference in mutation rates in the two groups on the partition $\calT_p^{(l)}$. 
		\item $H_{A,p}^{(l)}$: at least one of the $K+1$ sub-interval of $\calT_p^{(l)}$ has a difference in the mutation rates across the groups.
	\end{itemize}
	The posterior probabilities for the global null can be obtained as a function of the marginal likelihoods obtained from the hierarchical model above, e.g., the posterior probability of global association is the sum of the posterior probabilities of all non-null models. For the present case, we assume independence across different sub-regions and simply reject the global null if $\prod_{i=1}^{K+1}(1-\omega_{p,i}^{(l)}) < (1-\xi)^{K+1}$. The steps of the method are outlined in Algorithm \ref{alg:example}.
	
	\begin{figure*}
		\centering
		\begin{minipage}[t]{0.48\textwidth}
			\centering
			\raisebox{-\height}{
				\begin{tikzpicture}
					% Define nodes
					\node[obs] (y1) {$y_1$};
					\node[obs,right=2cm of y1] (y2) {$y_2$};
					\node[latent, above=of y1] (T1) {$\mathbf{\lambda_1}$};
					\node[latent, above=of y2] (T2) {$\mathbf{\lambda_2}$};
					\node[latent, above=of T1, xshift=-0.75cm]  (P) {$\mathbf{P}$};
					\node[latent, left=1cm of P]  (A) {$\alpha,\beta, M$};
					\node[latent, above=of T2]  (pi) {$\pi$};
					\node[latent, above=of P]  (G) {$\mathbf{P_0}$};
					% Connect the nodes
					\edge {T1} {y1} ; %
					\edge {T2} {y2} ;
					\edge {P} {T1} ;
					\edge {P,pi} {T2} ;
					\edge {T1} {T2} ;
					\edge {A,G} {P} ;
					% Plates
					\plate {yt1} {(T1)(y1)} {$N$} ;
					\plate {yt2} {(T2)(y2)} {$N$} ;
					\plate {pt} {(P)(T1)} {$\infty$} ;
			\end{tikzpicture}}
			\vspace{-0.1cm}
			\caption{Bayesian Nonparametric Model for Testing Differences}
			\label{fig:model}
		\end{minipage}\hfill
		\begin{minipage}[t]{0.48\textwidth}
			\centering
			% \alglanguage{pseudocode}
			\renewcommand\figurename{Algorithm}
			\setcounter{figure}{\value{algorithm}}
			\small
			\caption{Multiscale NP-BayesTest}
			\label{alg:example}
			\begin{algorithmic}[1]
				\STATE {\bfseries Input:} Count vectors $\y_{1}, \y_{2}$, positions $J$, domain-set $\calT$, depth of the binary tree $L$, threshold $\xi$.
				\FOR{$l$ = 1 \bfseries{to} $L$}
				\STATE Sort all domains $T \in \calT^{l}$ by positions $J$. 
				\FOR{all intervals $T \in \calT^{l}$}
				\STATE Let $\y_j^{T}; j = 1,2 $ and $J^{T}$ be the counts and the position-vector restricted to $T$. 
				\STATE Calculate $\omega_i^{T} = P(\lambda_{i,1} \neq \lambda_{i,2} | \y_1,\y_2), \; i = 1,\ldots, |J^T|$.
				\STATE Calculate probability of global association $P_{H_0}(T) = \prod (1-\omega_i^{T})$.
				\IF{$p_{H_0}^{T} > \xi^{|\J^T|} (\mbox{global threshold})$} 
				\STATE Prune interval $T$.
				\STATE Update $\calT \Leftarrow \calT \setminus T$, $J \Leftarrow J \setminus \{J \in T\}$. 
				\ENDIF
				\ENDFOR
				\ENDFOR   
			\end{algorithmic}
		\end{minipage}
	\end{figure*}
	A crucial issue is the choice of partitions and the depth of resolution. In the absence of any prior information about the location of the hot and cold spots, one would partition the genome using a balanced binary tree with partitioning informed by the posterior probabilities. 
	
	\section{Theoretical Properties: Weak Consistency}\label{sec:theory}
	We show that our proposed Dirichlet process mixture leads to posterior consistency in estimating the marginal probability mass function $P_0$ of the rare variants, and the consistency will continue to hold even after combining the estimates over multiple intervals. It will be interesting to settle theoretically if a multiresolution DPM model will put most of its posterior mass on a class of sparse density functions for the rate parameters. Although a thorough investigation of the properties of the induced multiple testing procedure for comparing the true abundances across groups is beyond the scope of the current article, we conjecture that the multiple testing rule in \eqref{eq:rule} will enjoy asymptotic optimality properties similar to \cite{bogdan2011asymptotic}. 
	
	\subsection{Dirichlet mixture of Poisson}
	Let $p_\lambda$ denote the Poisson probability mass function with mean $\lambda$. Let $\bTheta = \mathbb{R}^+$, and $\mathcal{M}$ be the set of probability measures on $\mathbb{R}^+$. For $P \in \mathcal{M}$, $f_P$ denotes the density:
	$$
	f_P(i) = \int_{0}^{\infty} p_\lambda(i) dP(\lambda) , i \in \mathbb{N}
	$$
	Let $\mathcal{F}_{\mathbb{N}}$ be the set of all probability mass functions in $\mathbb{N}$. Then the prior $\Pi$ on $\mathcal{M}$ induces a prior on $\mathcal{F}_{\mathbb{N}}$ through the map $f_P = \int p_\lambda dP(\lambda)$, which we continue to denote by $\Pi$. Our model can be written as $P \sim \Pi$ and given $P$, $X_1, \ldots, X_n \sim f_P$. Theorem \ref{th:kl} gives conditions for a class of densities to be in the K-L support of $\Pi$ and hence weakly consistent. The proof is similar to Theorem 3 in \citet{ghosal1999posterior} where they prove weak consistency for location mixtures of Gaussians for density estimation. 
	
	\begin{theorem}\label{th:kl}
		Let the true density $f_0 \equiv f_{P_0}$ be of the form $f_{P_0}(j) = \int p_\lambda(j) dP_0(\lambda)$ for $j = 1, 2, \ldots$ and let $K_\epsilon(f_0)$ be the $\epsilon$-weak neighborhood of $f_0$. If $P_0$ is compactly supported and belongs to the support of $\Pi$ , then $\Pi(K_\epsilon(f_0)) >0$ for all $\epsilon > 0$. 
	\end{theorem}
	This proves the weak consistency for the case where the true mass function is a Poisson or a mixture of Poissons over a compact set. We present the proof below.
	\begin{proof}
		Suppose $P_0[l,u] = 1$ for some $u>l>0$. This implies, $\Pi\{P: P[l,u] > \half \} >0,$ since $P_0$ is in the weak support of $\Pi$. 
		For $\eta > 0$, choose $k$ such that $\sum_{j>k} \mathrm{max} (1,j) f_0(j) < \eta$. We write: 
		\beq
		\sum_{j=0}^{\infty} f_{P_0}(j) \log \left( \frac{f_{P_0}(j)}{f_P(j)} \right) = \sum_{j \leq k} f_0 \log \frac{f_{P_0}}{f_P} + \sum_{j > k} f_0 \log \frac{f_{P_0}}{f_P}. \nonumber
		\eeq
		Consider the second term: 
		\begin{align*}
			\sum_{j > k} f_0 \log \frac{f_{P_0}}{f_P} & \leq \sum_{j >k} f_0(j) \log \left( \frac{\int_{l}^{u} p_{\lambda}(j) dP_0(\lambda)}{\int_{l}^{u} p_{\lambda}(j) dP(\lambda)} \right) \\
			& \leq \sum_{j > k} f_0(j) \log \left( \frac{\int_{l}^{u} \lambda^j e^{-\lambda} dP_0(\lambda)}{\int_{l}^{u} \lambda^j e^{-\lambda} dP(\lambda)} \right) \leq \sum_{j > k} f_0(j) \log \left( \frac{u^j e^{-l}}{l^j e^{-u}\half} \right) \\
			& = \sum_{j > k} f_0(j) \{ j \log(u/l) + (u-l) + \log 2 \}  \leq \eta \{ \log(u/l)+(u-l) + \log 2\}.
		\end{align*}
		Now consider the first term $\sum_{j \leq k} f_0 \log \frac{f_{P_0}}{f_P}.$
		Note that $\inf_{0\leq j \leq k} \inf_{l \leq \lambda \leq u} p_{\lambda}(j) = c >0$. Let 
		\begin{align*}
			E & = \left\{ P: \left|\int p_{\lambda}(j)dP(\lambda) - \int p_\lambda(j) dP_0(\lambda) \right| < c\delta, \; j \leq k \right\}.
		\end{align*}
		Since $E$ is a weak neighbourhood of $P_0$, $\Pi(E) > 0$. Take $P \in E$. For such a P, 
		\beq
		\left|\int p_{\lambda}(j)dP(\lambda) - \int p_\lambda(j) dP_0(\lambda) \right| < c \delta, \nonumber
		\eeq
		and hence, 
		$$
		\left|\frac{\int p_{\lambda}(j)dP(\lambda)}{\int p_\lambda(j) dP_0(\lambda)} - 1 \right| < \delta \Rightarrow \left|\frac{\int p_{\lambda}(j)dP_0(\lambda)}{\int p_\lambda(j) dP(\lambda)} - 1 \right| < \frac{\delta}{1-\delta}.
		$$
		Thus, we have proved that for $P$ in a set of positive $\Pi$-probability, 
		\begin{align*}
			\sum_{j=0}^{\infty} f_{P_0}(j) \log\left( \frac{f_P(0)(j)}{f_P(j)} \right) & \leq \frac{\delta}{1-\delta} + \eta \{ \log(u/l)+(u-l) + \log 2\}.
		\end{align*}
		One can choose $\eta$ and $\delta$ so that for any $u > l > 0$ the RHS is less than $\epsilon$. This completes the proof.
	\end{proof} 
	\textbf{Consistency over multiple resolutions} Consider the balanced binary tree in Algorithm \ref{alg:example}, where at level $l$ we partition the interval $\calT$ in $2^l$ intervals, denoted by $\calT_p^{(l)}$. We model the counts $y_p^{(l)}$ belonging to the interval ${\calT}_p^{(l)}$ as a Dirichlet Mixture of Poisson which possess weak consistency, and now we show that the weak consistency holds even when we combine estimates of probability mass functions (PMF) across partition to obtain an estimate of the true PMF for the entire interval $T$, for any resolution $l$. To see this, denote the aggregated PMFs as $f_0^{(l)} = \sum_p f_{0,p}^{(l)}$ and $f^{(l)} = \sum_p f_{p}^{(l)}$. By the weak consistency of DPM-Poisson, for any $\epsilon > 0$, 
	\beq
	\sum_{j} f_{0,p}^{(l)}(j) \log \frac{f_{0,p}^{(l)}(j)}{f_{p}^{(l)}(j)} < \epsilon, \; \mathrm{for\, all} \; l = 1, \ldots, L \mbox{ and } p = 1, \ldots, 2^l. \nonumber
	\eeq
	We want to show that the weak consistency holds even when we aggregate over the partitions, that is, for all $l = 1, \ldots, L,$ and for all $\epsilon > 0$,
	\begin{align*}
		\sum_j f_{0}^{(l)}(j) \log \frac{f_{0}^{(l)}(j)}{f^{(l)}(j)} & = \sum_{j} \sum_{p=1}^{2^l} f_{0,p}^{(l)}(j) \log \frac{\sum_{p=1}^{2^l} f_{0,p}^{(l)}(j)}{\sum_{p=1}^{2^l} f_{p}^{(l)}(j)} < \epsilon.
	\end{align*}
	We first state the following lemma: a well known result about convexity of K-L divergence: 
	\begin{lemma}
		The K-L distance $D(f||g)$ is convex in the pair $(f,g)$, i.e. if $(f_1,g_1)$ and $(f_2,g_2)$ are two pairs of PMFs, 
		\begin{align*}
			D & \left(\alpha f_1 + (1-\alpha) f_2 || \alpha g_1 + (1-\alpha) g_2 \right) \leq \alpha D \left(f_1 || g_1 \right) + (1-\alpha) D \left(f_2 || g_2 \right).
		\end{align*}
		This property can be generalized to $P$ pairs $(f_p, g_p),\, p = 1, \ldots, P$ and weights $\omega_p \geq 0, p = 1, \ldots, P$, $\sum \omega_p = 1$, i.e.,
		$$
		D \left(\sum \omega_p f_p || \sum \omega_p g_p \right) \leq \sum \omega_p D (f_p || g_p) 
		$$
	\end{lemma} 
	Taking $f_p = f_{0,p}^{(l)}, g_p = f_{p}^{(l)}$ and $\omega_p = 2^{-l}$ for $ p = 1, \ldots, 2^l$, it follows from the above lemma that: 
	\beq
	D(\sum_{p=1}^{2^l} f_{0,p}^{(l)} || \sum_{p=1}^{2^l} f_{p}^{(l)} ) \leq \sum_{p=1}^{2^l} D(f_{0,p}^{(l)} ||f_{p}^{(l)} ) \nonumber
	\eeq
	The result follows by noting that with positive $\Pi$-probability, $D(f_{0,p}^{(l)} ||f_{p}^{(l)} ) < \epsilon/2^l$ for all $p = 1, \ldots, 2^l$. 
	%%%%%%%%%%%%%%%%%%%%%%%%%%%%%%%%%%%%%%%%%%%%%%%%%%%%%%%%%%%

	\section{Numerical Experiments}\label{sec:sim}
	
	To illustrate the performance of the induced decision rule \eqref{eq:rule}, we consider the problem of testing for differences between realizations of two inhomogeneous Poisson processes having intensity functions:
	\begin{align*}
		\lambda_1(x) & = 2\exp \left(-\{(x-50)/10 \}^2 \right) +20\exp\left(-\{(x-10)/10\}^2 \right); x \in \mathbb{R}^+ \\
		\lambda_2(x) & = 20\exp \left(-\{(x-50)/10 \}^2 \right) +2\exp\left(-\{(x-10)/10\}^2 \right); x \in \mathbb{R}^+,
	\end{align*} 
	designed to have peaks at different locations $x = 10$ and $x = 50$ of different heights. We generate the count datasets $Y_j = \{y_{i,j}; i = 1, \ldots, m = 100\}, j = 1,2 $ as the number of events in a regular grid with $K$ equispaced points on $[0,70]$, with different resolutions $K = 35$, $K = 50$ and $K = 100$ to illustrate the performance at different scales.

	\begin{figure}[ht!]
		%\vskip 0.2in
		\begin{subfigure}[t]{0.33\linewidth}
			\begin{center}
				\centering
				\includegraphics[width=0.9\linewidth,height=2.5in]{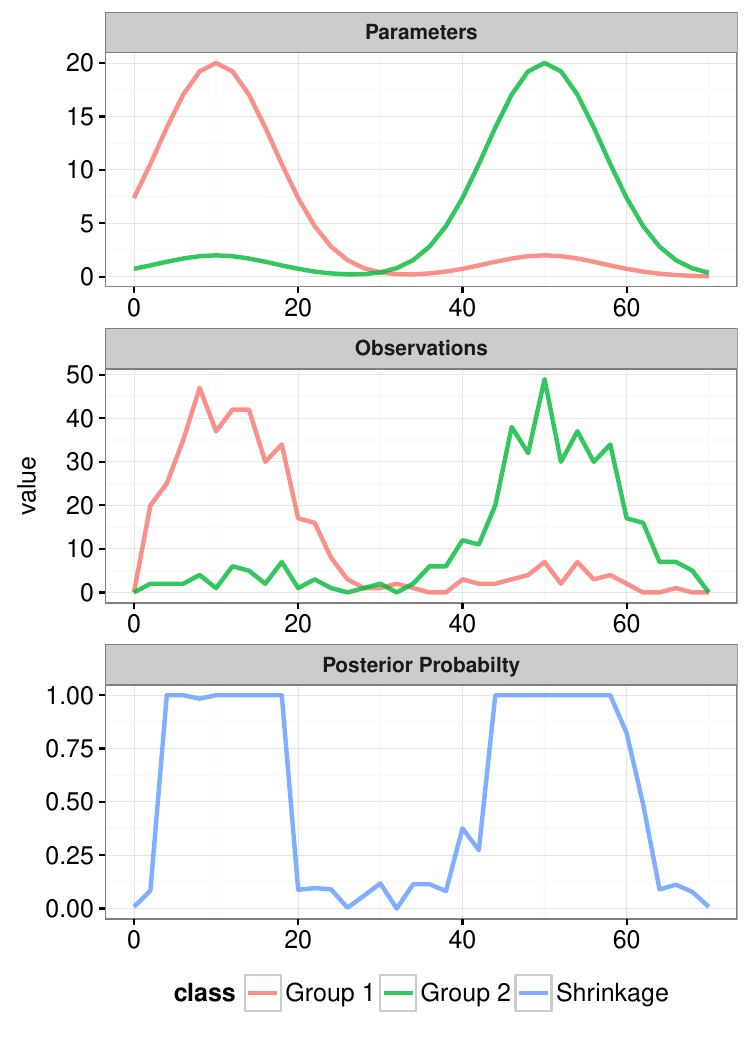}
				%\caption{The posterior probabilities of difference in Poisson rate parameters $\omega_i = P(\lambda_{i,1} \neq \lambda_{i,2} | \mbox{data})$ along with true intensity functions and the observations at multiple resolutions. }
				\caption{$35$ equispaced points.}
				\label{fig:sim1}
			\end{center}
		\end{subfigure}
		\begin{subfigure}[t]{0.33\linewidth}
			\centering
			\includegraphics[width=0.9\linewidth,height=2.5in]{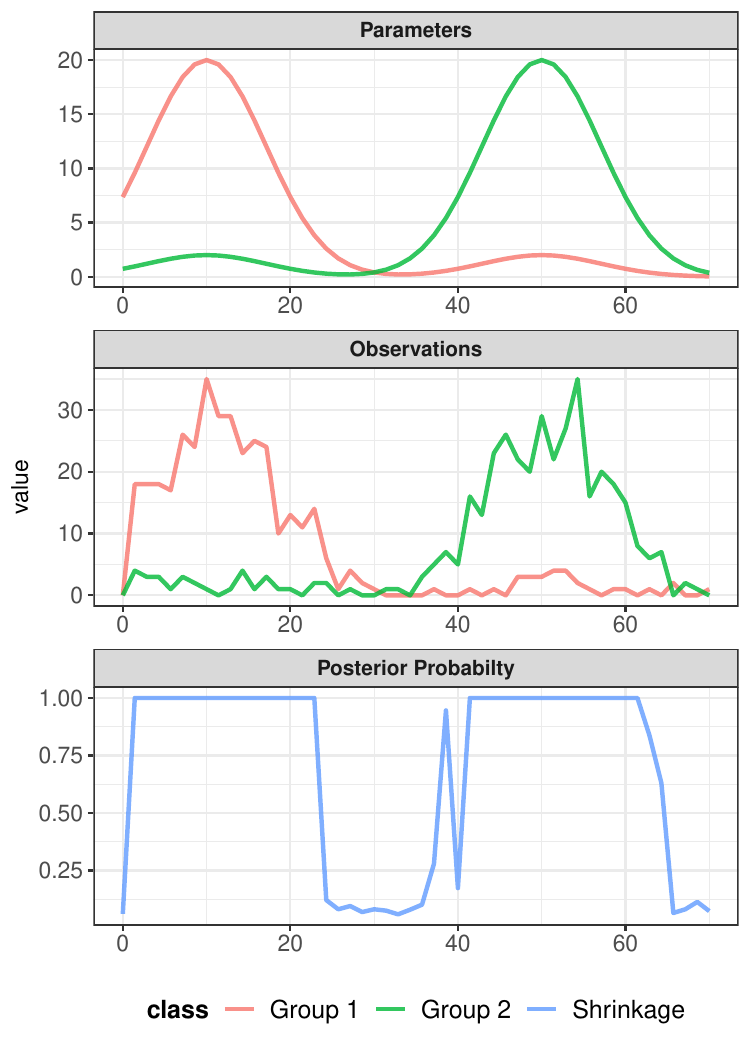}
			\caption{$50$ equispaced points.}
			\label{fig:sim2}
		\end{subfigure}
		\begin{subfigure}[t]{0.33\linewidth}
			\centering
			\includegraphics[width=0.9\linewidth,height=2.5in]{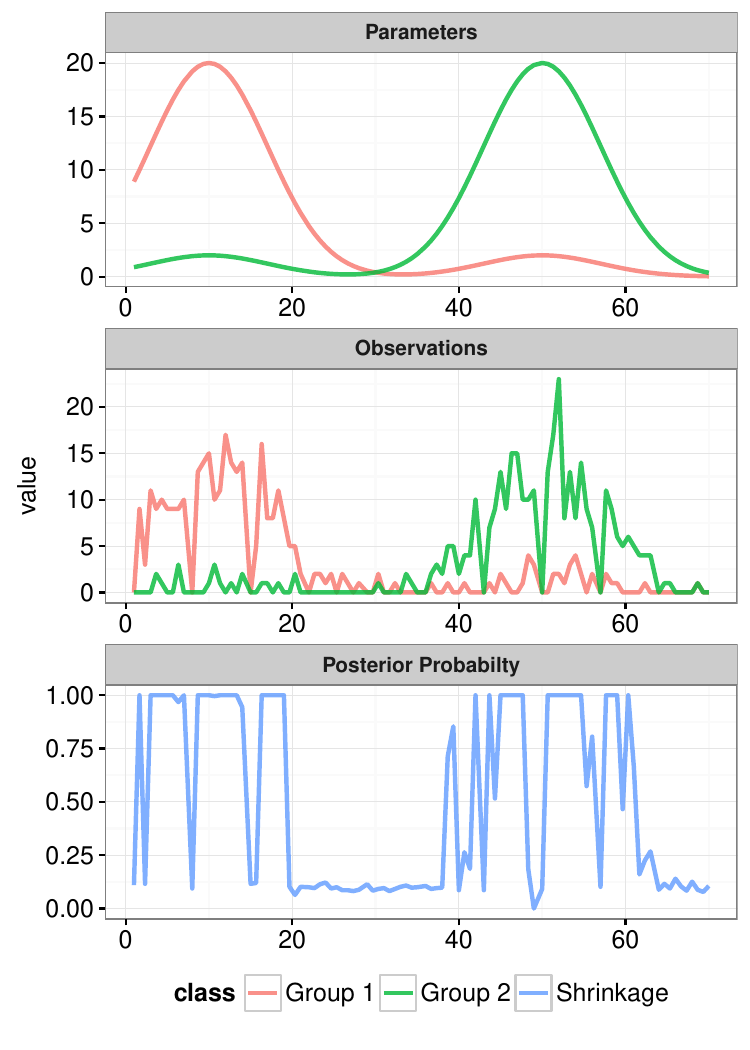}
			\caption{$100$ equispaced points.}
			\label{fig:sim3}
		\end{subfigure}
		\caption{The posterior probabilities of difference in Poisson rate parameters $\omega_i = P(\lambda_{i,1} \neq \lambda_{i,2} | \mbox{data})$ along with true intensity functions and the observations generated. }
		%\vskip -0.2in
	\end{figure} 
	
	Fig. \ref{fig:sim1} -- \ref{fig:sim3} show the posterior probabilities of differential abundance $\omega_i$ along with the true intensity functions $\lambda_{1}(\cdot), \lambda_2(\cdot)$ and the counts $\y_{j}$s for the two partitions considered. The figures suggest that the posterior probabilities for differential abundance are close to $1$ near both the peaks where the intensity functions differ significantly, i.e. both between $[0,20]$ and $[40,60]$. Fig. \ref{fig:sim2} and Fig. \ref{fig:sim3} illustrate the behavior at finer scales by recovering sub-intervals of differential intensity at the cost of increased computation time. To gain efficiency, we can safely prune the region where the intensity functions are similar, making the posterior probability close to $0$, e.g. the interval $[20,35]$ as it cannot contain a differential sub-region at a finer scale. The CPU time (\texttt{elapsed} time in \textsc{R}) for the three resolutions were 26.47, 27.03, and 40.56 seconds, respectively, on a Dell Latitude 5310 with an Intel(R) Core(TM) i7-10810U CPU @ 1.10GHz. The \textsc{R} codes for replicating this simulation experiment can be found at \texttt{\url{https://github.com/DattaHub/twosampleDPM}}. 
	
\begin{figure}[ht!]
		\centering
		\includegraphics[height=2.5in]{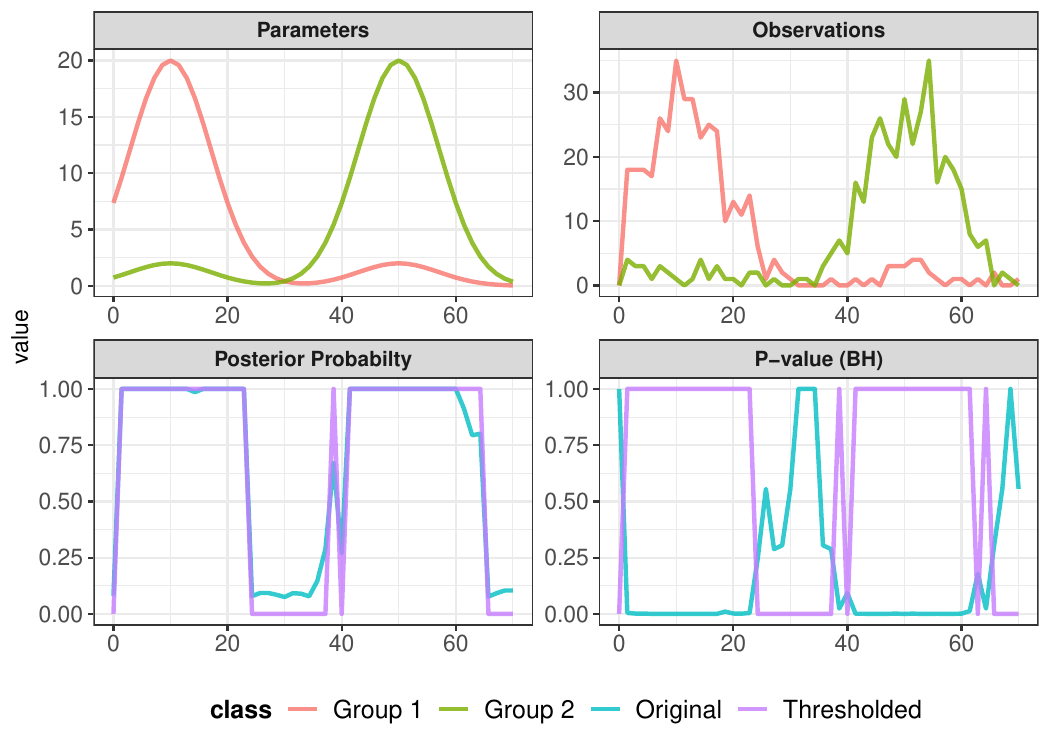}
		\caption{Comparison of pointwise exact test for two Poisson rates with the Benjamini--Hochberg FDR adjustment and the posterior inclusion probability obtained from the DPM-Poisson model in Fig. \ref{fig:model}.}
		\label{fig:compare-fdr}
	\end{figure}

	Although there does not exist a method that can be compared directly with our multiresolution-DPM model--either because they lack the multiresolution framework or the flexible nonparametric modeling part-- it is possible to compare the DPM-Poisson model with a standard likelihood-ratio based test after suitable adjustment. To compare, we apply the \texttt{poisson.test} in \textsc{R} to the two count series obtained at $K = 50$ resolutions and apply the Benjamini--Hochberg false discovery rate (FDR) control \citep{benjamini1995controlling}, and apply the usual $FDR \le 0.05$ cut-off for to the adjusted p-values. The positions of differential abundance thus identified are then compared with the posterior inclusion probabilities (PIP) along with a $0.5$ cut-off, corresponding to the median probability model \citep{barbieri2004optimal,barbieri2021median}. Figure \ref{fig:compare-fdr} shows the PIP versus p-values with FDR adjustment with overlaid binary series with the respective cut-offs: the methods agree on most parts, but the p-value based procedure leads to a missed discovery in the position $62.86$, with an adjusted p-value $0.0178$ and PIP of $0.794$. The results seem to suggest that the posterior inclusion probabilities $P(\lambda_{i,1} \neq \lambda_{i,2} | \y_1,\y_2)$ can be used for effectively controlling false discoveries without any further \textit{post hoc} adjustment like the p-value based procedures, and the DPM-Poisson model is expected to be robust to the data-generating mechanism unlike the likelihood-ratio based tests. We would also like to note here that an analogous frequentist test, parametric or nonparametric, with such a multi-resolution structure could be developed that has such desirable properties. 
	
	%%%%%%%%%%%%%%%%%%%%%%%5
	\section{Application to Rare Variants Data}\label{sec:real}
	
	We apply our method to assess differences between the mutation rates for cases and controls based on rare variants data 
	associated with common variable immunodeficiency ({\texttt{CVID}}). The background mutation rate for the control group is obtained from the Exome aggregation consortium ({\small ExAC}) that reports the total number of mutated alleles along the whole exome for 60,076 individuals. The observations represent rare variants ordered by their amino acid positions. We apply our method to the whole dataset and prune branches at each level based on the threshold for the probability of global null hypothesis restricted to each branch, and retain those that are likely to have at least one differentially mutated position. The successive levels are induced by the genes, the known protein domains within genes, and the amino acid positions within protein domains. If protein domain information is unavailable, one can grow the tree until sufficient data reduction has been achieved while keeping in mind that the probability of pruning a region with significant difference decreases with the number of levels of the tree. 
	
\begin{figure}[ht!]
		%\vskip 0.2in
		\begin{subfigure}[t]{0.5\linewidth}
			\centering
			\includegraphics[width=0.9\linewidth,height=2in]{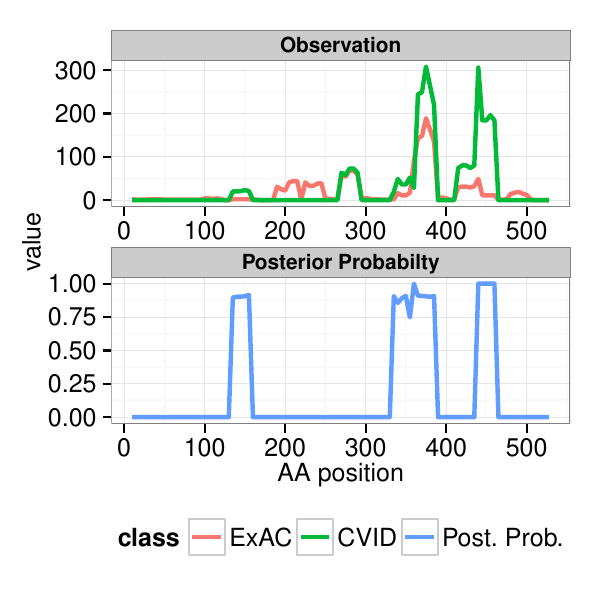}
			\caption{{\texttt{RAG2}}}
			%\caption{The performance of our method for detecting differences rare mutation rates between {\texttt{CVID}} cases and healthy controls on {\texttt{RAG2}} gene. The top and bottom pane shows the observed mutation data and the posterior probability of differential abundance respectively.}
			\label{fig:rag2}
		\end{subfigure}
		\begin{subfigure}[t]{0.5\linewidth}
			\centering
			\includegraphics[width=0.9\linewidth,height=2in]{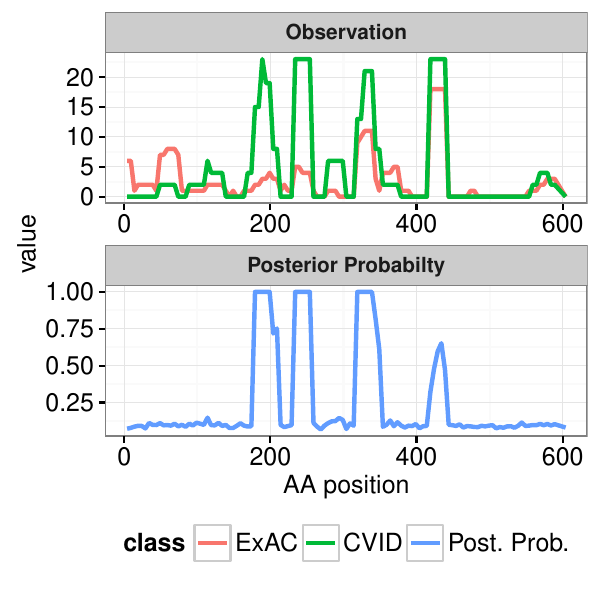}
			\caption{ {\texttt{SNCAIP}}}
			%\caption{The performance of our method for detecting differences rare mutation rates between {\texttt{CVID}} cases and healthy controls on {\texttt{RAG2}} gene. The top and bottom pane shows the observed mutation data and the posterior probability of differential abundance respectively.}
			\label{fig:sncaip}
		\end{subfigure}
		%\vskip -0.2in
		\caption{The performance of our method for detecting differences rare mutation rates between {\texttt{CVID}} cases and healthy controls on {\texttt{RAG2}} and {\texttt{SNCAIP}} gene. The top and bottom pane shows the observed mutation data and the posterior probability of differential abundance respectively.}
	\end{figure}

	We show the performance of our method with two protein-coding genes {\texttt{RAG2}} and {\texttt{SNCAIP}}. {\texttt{RAG2}} is involved in the development of B and T cells, and hyper-mutations on it have been associated with Omenn syndrome \citep{corneo2001identical}. The exome sequencing data for $n = 215$ {\texttt{CVID}} patients were collected in Dr. Sandeep Dave's lab at Duke Medicine. We consider the publicly available {\small ExAC} database as a control for detecting regions on the gene where the mutation rates are different. The ExAC database reports the total number of mutated alleles or variants along the whole exome for $60,076$ healthy individuals, and provides information about genetic variation in the human population, and acts as a reliable source of background mutation rate. The number of mutated alleles for the {\texttt{CVID}} and the {\small ExAC} control group at the $i^{th}$ position are denoted by $Y_{i,1}$ and $Y_{i,2}$ respectively. We model $Y_{i,1} \sim \mbox{Poi}(N_{i1}\lambda_{i,1})$ and  $Y_{i,2} \sim \mbox{Poi}(N_{i2}\lambda_{i,2})$ independently, where $\blambda_1$, $\blambda_2$ are the background mutation rates and $N_{i1}, N_{i2}$ are the total number of alleles at position $i$ for the cases and controls respectively. We can fix these as $N_{i1} = N = 2 \times 60,076=120,152$, and $N_{i2} = N = 2 \times 215 = 430$ for our problem, but in general, the number of total alleles varies depending on a lot of factors, including sequencing depth. The mutation frequencies and the probabilities for differential abundance are plotted in Fig. \ref{fig:rag2} and Fig. \ref{fig:sncaip}. 
	
	Although it is difficult to assess whether the detected regions of hyper-mutation are truly associated with {\texttt{CVID}} without further study, the results concur with our expectations: 1) the multi-resolution tree prunes regions where the mutation rate in {\texttt{CVID}} cases is not higher than the background, as well as 2) the detected differential regions on {\texttt{RAG2}} and {\texttt{SNCAIP}} fall in the known protein domain in the interval for amino acid positions. \footnote{For example, {\texttt{RAG2}} has a protein domain on $[50,389]$ and {\texttt{SNCAIP}} has a domain on  $[348,380]$ that are detected by our method.} Our method also outperforms the scan-statistics approach by \cite{ionita2012scan} on the same data set for detecting rare variants associated with {\texttt{CVID}}: the scan-statistics approach yields permutation P-values of $0.089$ and $0.034$ (based on $1,000$ permutations) for genes {\texttt{RAG2}} and {\texttt{SNCAIP}} respectively for testing association of variants residing in the known protein domain, leading to ambiguity in inference. The scan statistics take longer time as expected, registering 131.01 and 164.37 seconds for the two genes {\texttt{RAG2}} and {\texttt{SNCAIP}} for 1000 permutations, using the same computing resources as described in Section \ref{sec:sim}. In comparison, for our method, the CPU times recorded for the two genes {\texttt{RAG2}} and {\texttt{SNCAIP}} were 63.52 and 95.68 seconds, respectively.

	\section{Conclusive remarks}
	
	We introduce a new multi-scale test of the difference between the rate of incidences between two groups, motivated by the problem of comparing rare variants profiles across two disease subtypes. The multi-resolution framework helps us substantially reduce the computational burden by focusing the search away from regions of no difference, informed by testing a composite null hypothesis. We use a Dirichlet process mixture of Poisson with Gamma base measure for modeling the intensity of each process with a spike-and-slab prior on differences. There are several future directions for this methodology and theory. Firstly, in many application areas such as environmental criminology, spatial data are aggregated to a pre-determined grid-level resolution for both response and predictor variables \citep{ek2023quantifying}. A common feature in many of the models is that they combine geographic features such as built environments and socioeconomic variables at a chosen grid level over a geography. For example, the popular risk terrain modeling (RTM) \citep{caplan2011risk, caplan2015risk} creates a separate map layer for each predictor using a lattice grid of polygons, that are then combined to produce a composite map where the contribution or importance of each factor can be evaluated in a model-based way. Extending the multi-resolution Poisson-Gamma DPM for spatial data would be useful in such settings where spatial clustering or concentration of crime could aid in efficiency by the multi-resolution trimming approach. A second possible extension could be the use of shrinkage priors that could adapt to a broader range of sparsity, \textit{e.g.}, global-local priors for count data with quasi-sparsity such as the Gauss-hypergeometric \citep{datta2016bayesian} or extremely-heavy-tailed prior \citep{hamura2022global}. Finally, it would be interesting (from a purely theoretical viewpoint) to derive posterior contraction rates for the DPM-Poisson-Gamma model endowed with the two-groups test, similar to \citet{ghosal1999posterior}. Finally, as pointed out by a referee, we believe that an extension to more than two groups is plausible within this framework, perhaps using a recursive scheme, depending on the context. For example, in genomics application, one could think of a baseline group and multiple disease sub-types and extend our DPM-Poisson model accordingly. Such an extension would be interesting to build but is currently out of scope for this manuscript. 
	
	\section*{Data Availability Statement}
	The exome sequencing data for \texttt{CVID} patients were collected in Dr. Sandeep Dave's lab at Duke Medicine. The data are secondary, retrospective, and completely de-identified and are available from the authors on request.
	%% Acknowledgements should only appear in the accepted version. 
	\section*{Acknowledgements} 
	The authors would like to thank the AE and two anonymous referees whose comments led to substantial improvements in the revised version of the manuscript. Sayantan Banerjee is supported by SERB MATRICS Grant MTR/2022/000714, Govt. of India.
	%
	%
	%% In the unusual situation where you want a paper to appear in the
	%% references without citing it in the main text, use \nocite
	%\nocite{langley00}
	
	%\small
	%\bibliographystyle{agsm}
	\bibliographystyle{apalike}
	\bibliography{icmlref}
	
\end{document}